\documentclass[conference]{IEEEtran}
\IEEEoverridecommandlockouts
\usepackage{cite}
\usepackage{amsmath,amssymb,amsfonts}
\usepackage{algorithmic}
\usepackage{graphicx}
\usepackage{textcomp}
\usepackage{xcolor}
\usepackage{hyperref}
\usepackage{paralist}
\usepackage{xspace}
\usepackage{booktabs}
\usepackage{enumitem}

\makeatletter 
\newcommand{\linebreakand}{%
  \end{@IEEEauthorhalign}
  \hfill\mbox{}\par
  \mbox{}\hfill\begin{@IEEEauthorhalign}
}
\makeatother

\def\BibTeX{{\rm B\kern-.05em{\sc i\kern-.025em b}\kern-.08em
    T\kern-.1667em\lower.7ex\hbox{E}\kern-.125emX}}

\def\somon{\emph{SOMONITOR}\xspace}
\def\grab{\emph{Grab}\xspace}
\def\gojek{\emph{Gojek}\xspace}
\def\sowide{\emph{SoWide-v2}\xspace}

\begin{document}

\title{SOMONITOR: Combining Explainable AI \& Large Language Models for Marketing Analytics}

\author{
\IEEEauthorblockN{Aleksandr Farseev}
\IEEEauthorblockA{
\textit{SoMin.ai Research} \\
Singapore, Singapore \\
sasha@somin.ai}
\and
\IEEEauthorblockN{Qi Yang}
\IEEEauthorblockA{
\textit{SoMin.ai Research} \\
Singapore, Singapore \\
yang@somin.ai}
\and
\IEEEauthorblockN{Marlo Ongpin}
\IEEEauthorblockA{
\textit{SoMin.ai Research} \\
Kuala Lumpur, Malaysia \\
marlo@somin.ai} 
\and
\linebreakand 
\IEEEauthorblockN{Ilia Gossoudarev}
\IEEEauthorblockA{
\textit{ITMO University} \\
Saint Petersburg, Russia \\
goss@itmo.ru}
\and
\IEEEauthorblockN{Yu-Yi Chu-Farseeva}
\IEEEauthorblockA{
\textit{SoMin.ai Research} \\
Singapore, Singapore \\
joy@somin.ai}
\and
\IEEEauthorblockN{Sergey Nikolenko}
\IEEEauthorblockA{
\textit{ITMO University,} \\
\textit{St. Petersburg Department of the Steklov} \\ \textit{Institute of Mathematics}, Saint Petersburg, Russia \\
sergey@logic.pdmi.ras.ru}
}

\maketitle

\begin{figure*}\centering
  \includegraphics[width=.95\textwidth]{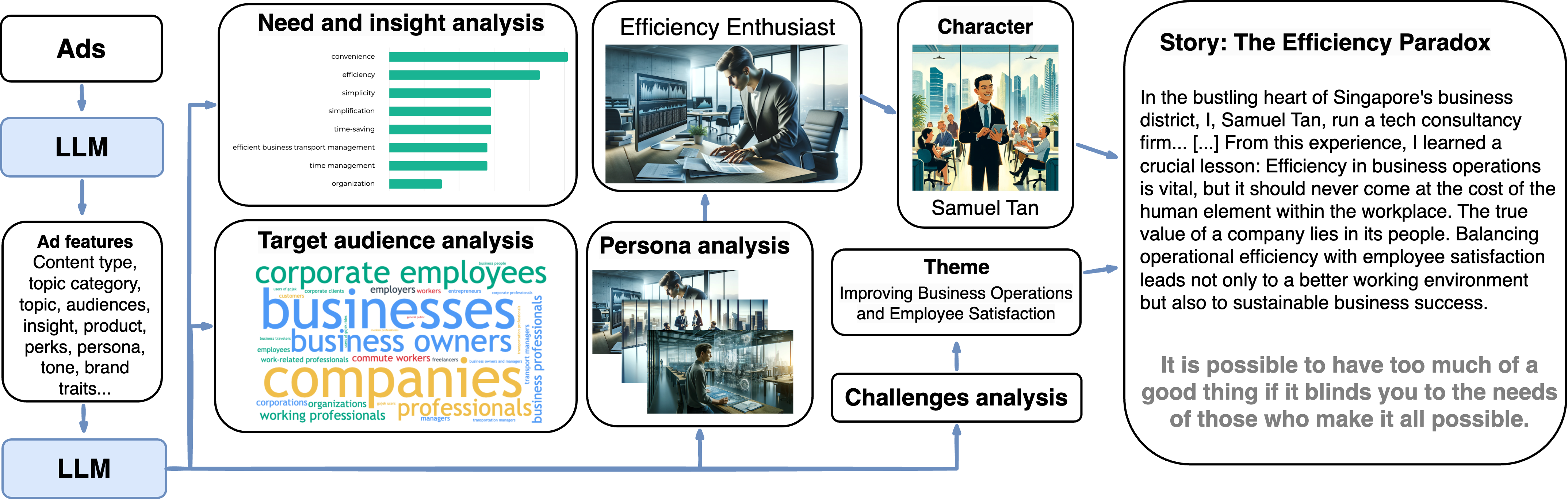}
  \caption{General pipeline of the \somon advertising analysis framework.}
  \label{fig:teaser}
\end{figure*}

\begin{abstract}
Online marketing faces formidable challenges in managing and interpreting immense volumes of data necessary for competitor analysis, content research, and strategic branding. It is impossible to review hundreds to thousands of transient online content items by hand, and partial analysis often leads to suboptimal outcomes and poorly performing campaigns.
We introduce an explainable AI framework \somon that aims to synergize human intuition with AI-based efficiency, helping marketers across all stages of the marketing funnel, from strategic planning to content creation and campaign execution. \somon incorporates a CTR prediction and ranking model for advertising content and uses large language models (LLMs) to process high-performing competitor content, identifying core content pillars such as target audiences, customer needs, and product features. These pillars are then organized into broader categories, including communication themes and targeted customer personas.
By integrating these insights with data from the brand's own advertising campaigns, \somon constructs a narrative for addressing new customer personas and simultaneously generates detailed content briefs in the form of user stories that, as shown in the conducted case study, can be directly applied by marketing teams to streamline content production and campaign execution.
The adoption of \somon in daily operations allows digital marketers to quickly parse through extensive datasets, offering actionable insights that significantly enhance campaign effectiveness and overall job satisfaction.
\end{abstract}

\begin{IEEEkeywords}
Digital Advertising, Ads Performance Prediction, Deep Learning, Large Language Model, Explainable AI
\end{IEEEkeywords}

\section{Introduction}

The contemporary online marketing landscape is bursting with an overwhelming amount of content in the form of databases of advertisements, marketing campaigns of various competitors, properties of user audiences, and more. Marketers must not only sift through this information but also extract meaningful insights regarding audience segmentation, trending topics, and customer needs. 
Recent advances in the field of large language models (LLMs) have made them into tools of choice in various domains (see Section~\ref{sec:related}). These models, trained on huge datasets, exhibit remarkable capabilities in both content analysis and content generation. In online marketing, 
modern LLMs are already able to offer a promising solution by processing and analyzing large datasets to identify key elements, facilitating a more targeted and effective marketing strategy. Moreover, they can help generate new advertisement content, both text-based and multimodal.

Despite all this progress, in this work we note that vanilla LLMs also have limitations in \emph{content scoring}, which is a crucial aspect of effective marketing. We show that 
a small neural network, specifically tailored for CTR prediction based on advertising content, performs better at content scoring than standard LLMs, even those enhanced with multimodal capabilities, offering a more reliable tool for marketers. 

Therefore, we propose a system, called \somon, that combines predictive analytics and LLMs in a unified online marketing assistant. Importantly, marketing strategies encompass more than just data analysis; they involve understanding and implementing effective communication styles. It is essential to utilize these communication styles not merely as classification labels but as dynamic personas. This approach enables the identification of different communication schemas tailored to diverse user personas. These schemas, and the user personas themselves, can be analyzed in existing marketing campaigns via a {clustering} procedure (Section~\ref{sec:perstheme}).
The next step involves aligning these personas with appropriate communication styles to craft compelling narratives. These narratives, or \emph{stories}, are essential content that marketers can directly employ in their campaigns for brief production or content calendar building. However, to construct these stories one has to use high-quality data, including marketing communications and advertisements that have previously proven to be successful in building rapport with the corresponding user personas.

Thus, the core issue circles back to the need for an advanced, yet explainable, predictive model for content scoring and improvement, presented in Section~\ref{sec:ctr}. This model can be used to filter and identify ``good'' advertisements that can then serve to enhance the efficiency of LLMs in generating impactful marketing content. Our research addresses this need by introducing a predictive model that refines data input for LLMs, ensuring the generation of content that resonates well with target audiences and supports strategic marketing initiatives; in Section~\ref{sec:story}, we show how this model alone can be used for iterative content improvement via interactive heatmap generation but also fits together with an LLM in a unified workflow that can generate complete user stories that being used by real businesses drastically improve their productivity. The overall pipeline of \somon is illustrated in Fig.~\ref{fig:teaser}.


\section{Related work}\label{sec:related}

Large language models (LLMs) 
have been successful in numerous applications~\cite{forecast5030030,minaee2024large} and
a wide variety of tasks, from low-resource machine translation~\cite{enis2024llm} to writing source code for full-scale applications~\cite{chen2021evaluating,jiang2023mistral} and solving problems that require mathematics and engineering design~\cite{kevian2024capabilities}. 
Importantly, the quadratic complexity bottleneck inherent in the original Transformer architecture~\cite{DBLP:journals/corr/VaswaniSPUJGKP17} is also being overcome with sparse attention mechanisms~\cite{beltagy2020longformer,child2019generating}, low-rank decomposition of attention weight matrices~\cite{wang2020linformer,choromanski2022rethinking}, Mamba-style models~\cite{gu2024mamba,qu2024surveymamba,bansal2024comprehensivesurveymambaarchitectures}, and other modifications with linear complexity~\cite{ma2023mega,pmlr-v202-orvieto23a}. As a result, latest LLMs can extend their context windows (i.e., input sizes) up to millions of tokens~\cite{geminiteam2024gemini,claude3}, which makes it possible to use LLMs directly for mid- and large-scale analysis of text datasets.

Recently, LLMs have also obtained multimodal capabilities, allowing to use them for video understanding~\cite{tang2024video}. The recently released GPT-4o, where 'o' stands for ``Omni'', brings seamless multimodality of image, audio, and video analysis to state-of-the-art LLMs~\cite{gpt4o}. LLMs can be combined with diffusion-based generative models~\cite{NEURIPS2020_4c5bcfec,song2021denoising,Rombach_2022_CVPR} to achieve state-of-the-art image and video generation capabilities~\cite{ho2022imagen,hong2022cogvideo,liu2024sora}.

Moreover, LLMs are increasingly used to process structured data, both in order to clean and/or augment data for ML model training and to directly improve the user experience. Extensive datasets used to train LLMs themselves~\cite{kocetkov2023the,DBLP:journals/corr/abs-2101-00027,patel2020introduction,tokpanov2024zyda13tdatasetopen} require rigorous filtering and ensuring data quality, for both technical~\cite{DBLP:journals/corr/abs-2105-02732,DBLP:journals/corr/abs-2112-11446,elazar2024whats,lee-etal-2022-deduplicating} and legal considerations~\cite{min2024silo,pmlr-v202-yu23g,doi:10.1126/science.adi0656}. High-quality datasets are equally essential for LLM fine-tuning, particularly for instruction tuning, which depends on well-crafted full-text prompts and responses~\cite{10.5555/3618408.3619349,DatabricksBlog2023DollyV2,sanh2022multitask}, and LLMs themselves increasingly assist in improving dataset quality; for instance, CoachLM~\cite{10597991} autonomously revises samples in instruction tuning datasets, IterClean~\cite{10.1145/3674399.3674436} introduces an LLM-driven framework for cleaning data, and the Natural Instructions dataset~\cite{wang2022supernaturalinstructionsgeneralizationdeclarativeinstructions} has been expanded with synthetic data in Unnatural Instructions~\cite{honovich-etal-2023-unnatural}. Other efforts include purifying datasets of copyright and poisoning issues with smaller LLMs~\cite{li2024purifyinglargelanguagemodels}, among other approaches~\cite{zhang2024instructiontuninglargelanguage}. 

LLMs are also employed as tools for dataset preprocessing and refinement; e.g., Jellyfish~\cite{zhang-etal-2024-jellyfish} specializes in data preprocessing, LLMClean~\cite{10.1007/978-3-031-70421-5_7} focuses on cleaning tabular data, CAAFE~\cite{hollmann2023large} generates meaningful features for tabular datasets, and other works~\cite{badaro-etal-2023-transformers,10.1145/3616855.3635752} explore tabular data processing, even showcasing how LLMs can synthesize realistic synthetic data~\cite{borisov2023language}. In the realm of databases and data lakes, LLMs have been applied to generate structured views~\cite{10.14778/3626292.3626294}, identify necessary data transformations~\cite{10597732}, and enable natural language access to relational databases~\cite{10598154,10597914,10184517,10184611,10184547}, e.g., for exploratory data analysis~\cite{10597681}. They also aid in interacting with graph databases~\cite{10597730}, answering questions over relational datasets~\cite{qin2024relationaldatabaseaugmentedlarge}, and other related tasks~\cite{10.14778/3685800.3685838,ZSL24}.

One important aspect of data processing with LLMs deals with recommender systems, where LLMs can unify tabular data processing and semantic understanding of items and context~\cite{10506571}. 
LLMs can be used directly for recommendation~\cite{liu2023chatgptgoodrecommenderpreliminary,chen2023knowledgegraphcompletionmodels}, especially in conversational systems where the used is engaged in a dialogue with the system~\cite{wang-etal-2023-rethinking-evaluation,He_2023,gao2023chatrecinteractiveexplainablellmsaugmented}, but usually an LLM-based recommender 
is either trained fully with a separate Transformer-based architecture---such systems include the P5 approach (Pretrain, Personalized Prompt \& Predict Paradigm)~\cite{10.1145/3523227.3546767,xu2024openp5opensourceplatformdeveloping}, M6-Rec~\cite{cui2022m6recgenerativepretrainedlanguage}, and IRS~\cite{10184767}---or fine-tuned with instruction tuning~\cite{zhang2023recommendationinstructionfollowinglarge}, as in RecLLM~\cite{friedman2023leveraginglargelanguagemodels}, LC-Rec~\cite{10597986}, TallREC~\cite{Bao_2023}, GLRec~\cite{wu2023exploringlargelanguagemodel}, or BERT-Trip~\cite{10184553}; 
text and Transformer-based embeddings of text can be used as user and item representations instead of ID embeddings~\cite{10598127,10.1145/3543507.3583434}. Multimodal LLMs have led to modality-based recommender systems that fuse encoders for different modalities~\cite{10.1145/3539618.3591932}, including sequence representations~\cite{10.1145/3534678.3539381}; multimodality can make recommenders more robust to domain transfer~\cite{wang2022transreclearningtransferablerecommendation,10597830,guo2023automatedpromptingnonoverlappingcrossdomain}.

Interpreting data through human-perceivable dynamic categories, such as topics or audiences, represents a significant step towards explainable AI. Nonetheless, the sheer volume of content and its prioritization continue to pose challenges, with limitations inherent in LLMs, in particular related to their hallucination phenomena~\cite{zhang2023siren}. We will see below that an LLM can say that the primary audience for 1000 pieces of marketing content is, e.g., ``families'', but without predictions of the content's popularity across this audience this insight is not actionable; LLMs are known to perform poorly and inconsistently in ranking problems~\cite{chen2023hallucination}. At the same time, online content platforms such as social networks and search engines, namely \emph{Facebook} (\emph{Meta})~\cite{fb-ctr-practical-lessons}, \emph{Google}~\cite{goog-ctr-view-from-trenches}, \emph{Alibaba}~\cite{baba-rep-learning-ctr}, and \emph{Taobao}~\cite{baba-image-matters}, have invested considerable efforts towards content prioritization and scoring. Unfortunately,
these results are based on private datasets, inaccessible to external content creators such as digital marketers. Proprietary engagement data used for content scoring models remains exclusive to the platforms, so independent creators cannot accurately predict or enhance the efficacy of their content. 

From all the above, it is evident that the comprehensive capabilities of LLMs position them as high-level assistants across a myriad of tasks within the realm of online marketing. However, there exist relatively few research papers devoted to these potential applications~\cite{feizi2024online}. 
The marketing industry is currently only starting to incorporate LLMs into the workflows, with many recent whitepapers and business surveys of possible applications, problems, and potential breadth of adoption of LLMs~\cite{TM23,KeB24}. 
Our recent work~\cite{10.1145/3581783.3612817} presented a comprehensive online marketing analysis system based on large language models; in this paper, we build upon this work and combine and further improve LLM-based analysis with predictive analytics. A key advantage noted in~\cite{10.1145/3581783.3612817} was the fact that LLM answers are understandable for humans and actionable in terms of business results: LLMs such as GPT-3.5~\cite{ouyang2022training} and GPT-4~\cite{openai2023gpt4} were already able to explain the ``reasoning'' behind recommender models and provide aggregate insights about advertising campaigns.
Before LLMs, text corpora in the context of recommender systems had been aggregated via topic modeling~\cite{10.1145/2615569.2615680,10.1007/978-3-319-27101-9_5}, sometimes also with deep learning~\cite{TN18} and user profiling~\cite{farseev2,TN17}, but topic modeling is based on the bag-of-words assumption and cannot summarize text as an LLM does. Visual understanding of ads has also been explored with convolutional networks~\cite{savchenko-etal-2020-ad}.

However, a robust technology that enables effective differentiation of content sources for insight extraction, as well as a cohesive framework for data exploration that integrates both content differentiation and LLM-based explanations, is still sorely needed. Our work aims to address these gaps by developing methods that enhance the functionality of LLMs in online marketing applications, facilitating more efficient utilization of modern AI in real-world scenarios.

\section{Content pillar identification}\label{sec:pillars}

\begin{figure}[!t]\centering
\includegraphics[width=\linewidth]{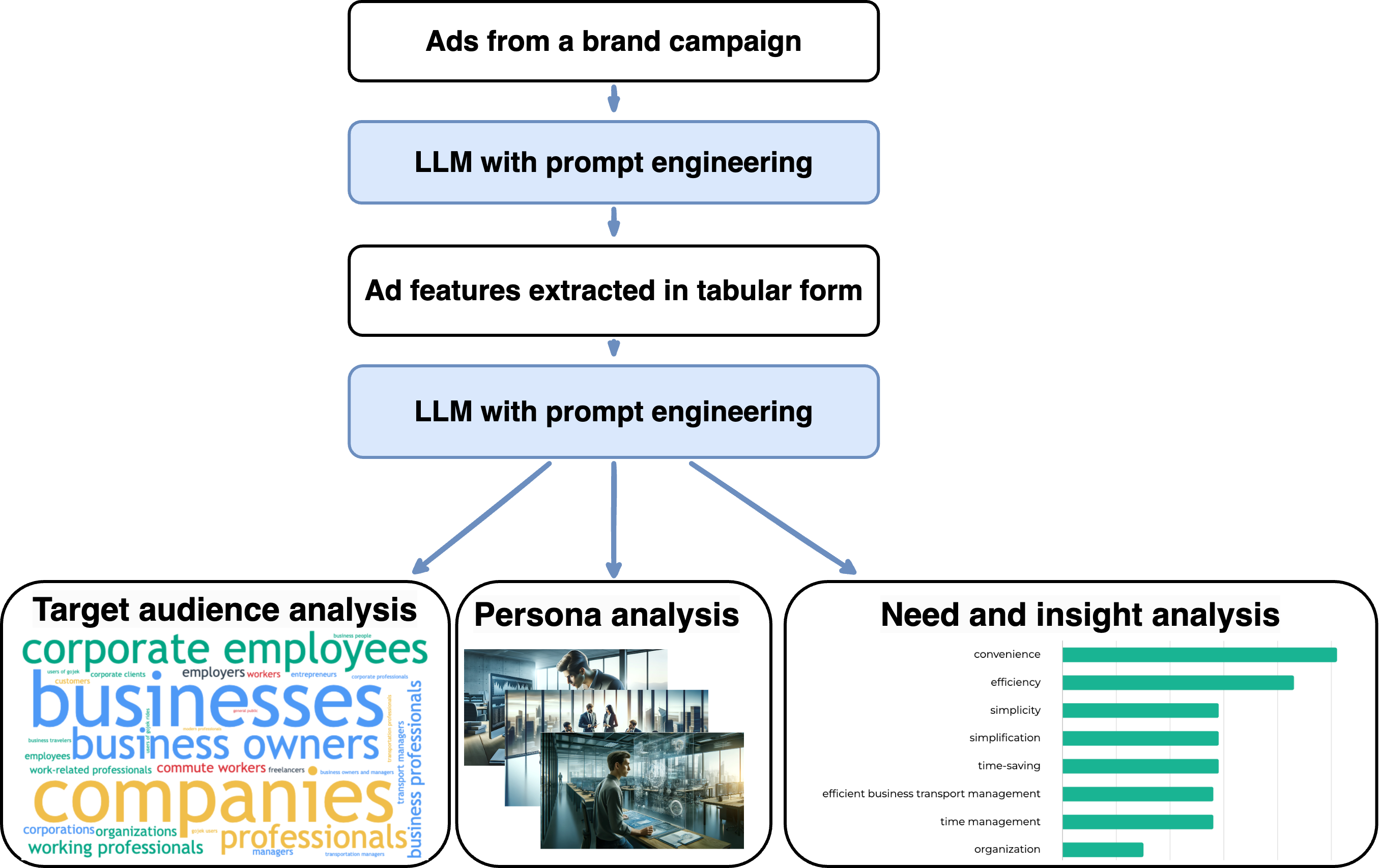} 

\caption{General pipeline of our LLM-based analysis}\label{fig:pipeline}
\end{figure}

As noted above, data handlers in digital marketing frequently encounter large volumes of creative assets to analyze, along with numerous ongoing promotional activities from competitors in an ``always-on'' environment. Given this context, conducting detailed analyses for each advertisement or even for every specific competitor campaign becomes impractical, highlighting the need for automated analytical tools. These tools are crucial for assisting marketers in quickly assimilating, synthesizing, and driving inferences from extensive data and information available to them.

To address this challenge, we propose to integrate a novel method into our framework that leverages LLMs to provide semantic insights into content, referred to as ``content pillars'' below. We outline an analytical pipeline that incorporates explanations and generations based on LLMs, demonstrating its practical utility through an empirical scenario involving four Singapore-based telecommunication companies. This component of our framework aims to enhance the interpretability and understanding of data, thereby supporting more informed decision-making in these dynamic and competitive sectors.

The architecture of our analytics is illustrated in Fig.~\ref{fig:pipeline}. Initially, we employ an LLM to extract specific, well-defined insights from the input content, such as the customer needs addressed by the advertisement and the products being advertised. These insights are then archived as \emph{content pillars} in a tabular format. Next, these pillars are processed with custom-crafted prompts to perform a generalized analysis of a brand's target demographics, personas, needs, and insights conveyed by the advertisements, as well as the tone and topical categories of both current and previous campaigns. This comprehensive analysis mirrors the campaign assessments typically conducted by marketing professionals and can be further used to refine a brand's messaging, tone, and target demographics. Below we explore these elements in detail.

\begin{figure}[!t]\centering
\includegraphics[width=\linewidth]{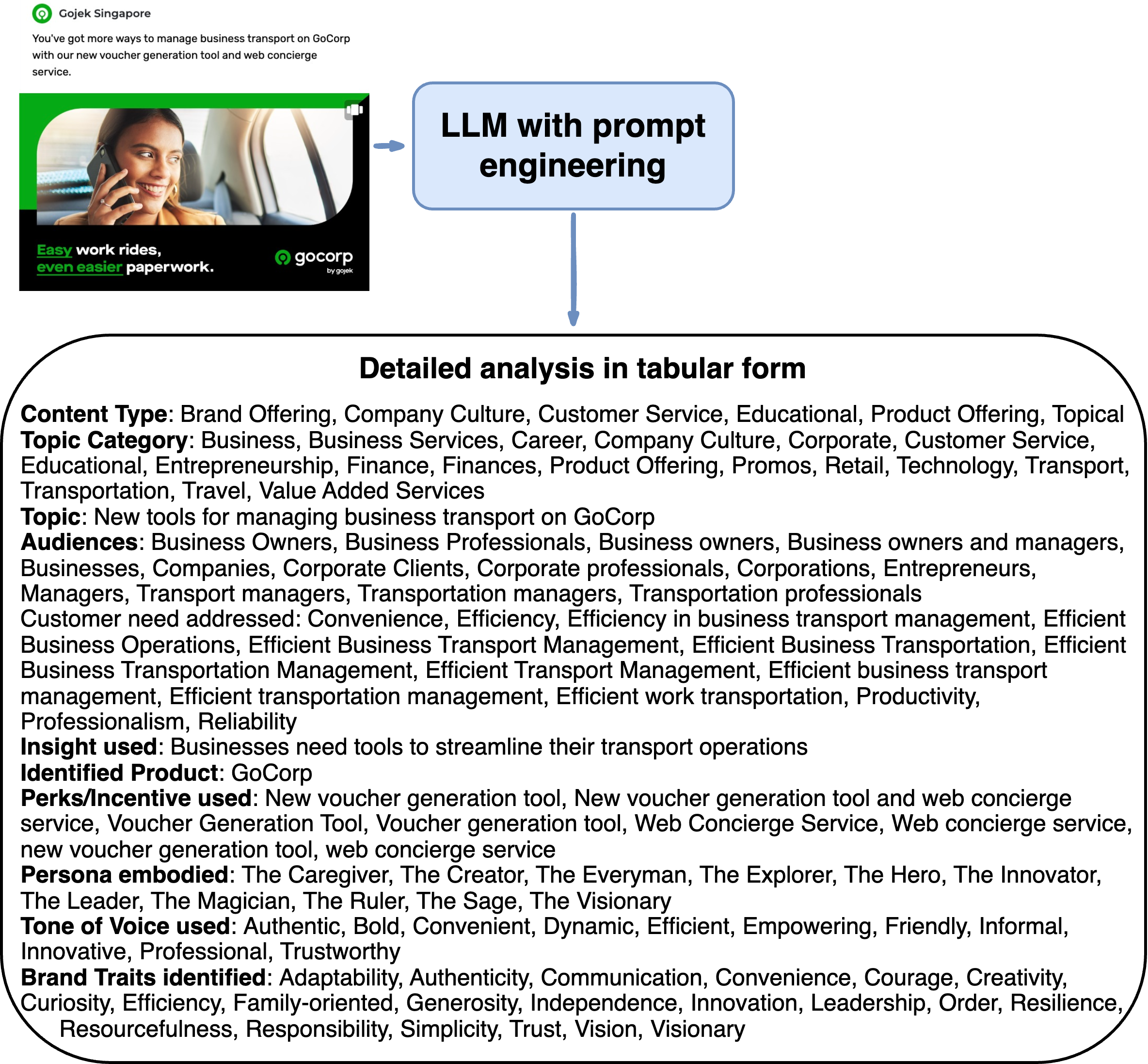} 

\caption{Sample ad analysis}\label{fig:adanalysis}
\end{figure}

Figure~\ref{fig:adanalysis} presents sample qualitative results of our experiments on the analysis of content pieces. We curated batches of advertisements from the \emph{Facebook Ad Library} for the same brand and processed them through an LLM with custom prompts. As a result, the LLM successfully identified key attributes of each advertisement, providing adequate responses to ostensibly ``human'' inquiries such as pinpointing the human need, insight, and predominant archetypes in a specific ad. Note that responses to most queries are standardized (as directed by the prompt) and amenable to automated processing. This analysis has always been important in online marketing but to our knowledge has not been automated and scaled before, always requiring human labor and thus being limited to a small sample of ads rather than an extensive dataset.

\section{Mining Communication Personas and Themes}\label{sec:perstheme}

\subsection{Problem setting}

In the field of marketing, utilizing explainable data descriptors such as the content pillars is essential. However, to significantly enhance a brand's marketing and communication strategies it is necessary to go beyond mere description and engage in inferential data analysis. Unfortunately, marketers often lack the necessary skills in data operations, hindering their ability to derive meaningful conclusions from the data, thereby requiring additional support.
To bridge this gap, we propose a unified framework \somon that unites LLM-based analysis with content scoring,
designed to provide marketers with higher-level content observations.

In marketing, the narrative constructed by any content emerges from an underlying story, which is a synthesis of a communication theme and a specific customer persona that the campaign aims to engage. We will come back to the stories in Section~\ref{sec:story}, illustrating this concept through a representative example, highlighting the integral relationship between the storytelling elements and the targeted persona. This approach empowers marketers to craft more resonant and effective communication strategies by aligning their messaging with the expectations and needs of their intended audiences. But to get to a good story, we need to extract two essential storytelling components: personas and challenges.

\subsection{Persona Identification}\label{sec:persona}

In marketing, the \emph{customer persona} is developed based on data about the audience of a brand, supported by other content pillars such as customer needs, interests, and aspirations~\cite{malik2019persona}. We have used X-Means clustering~\cite{pelleg2000x} to divide the content pieces into distinct customer personas based on brand content, specifically clustering ADA embeddings~\cite{neelakantan2022text} of the ``Audience'' content pillar as the main semantic differentiator.
X-Means was run with $3$ starting centroids and $k_{\max}=50$, identifying the number of clusters with BIC values~\cite{pelleg2000x}.

For this study, we analyzed data from the Gojek car-hailing brand and its main competitor in Singapore, \grab, focusing exclusively on the B2B business audience to limit the number of personas. The data for this experiment, collected over 2023 and 2024, comprised $5{,}967$ content pieces, with $3{,}703$ published as ads and $2{,}264$ as organic social media posts. Since most of the content related to B2B product offerings for both brands was presented in ads, we focused solely on the advertising content, totaling $1{,}120$ ads related to business products. Brand-wise, \gojek ran the majority of the ads ($76.8$\%, or $849$ ads), while \grab accounted for the remaining $23.2$\% ($271$ ads).

\begin{figure}[!t]
\includegraphics[width=\linewidth]{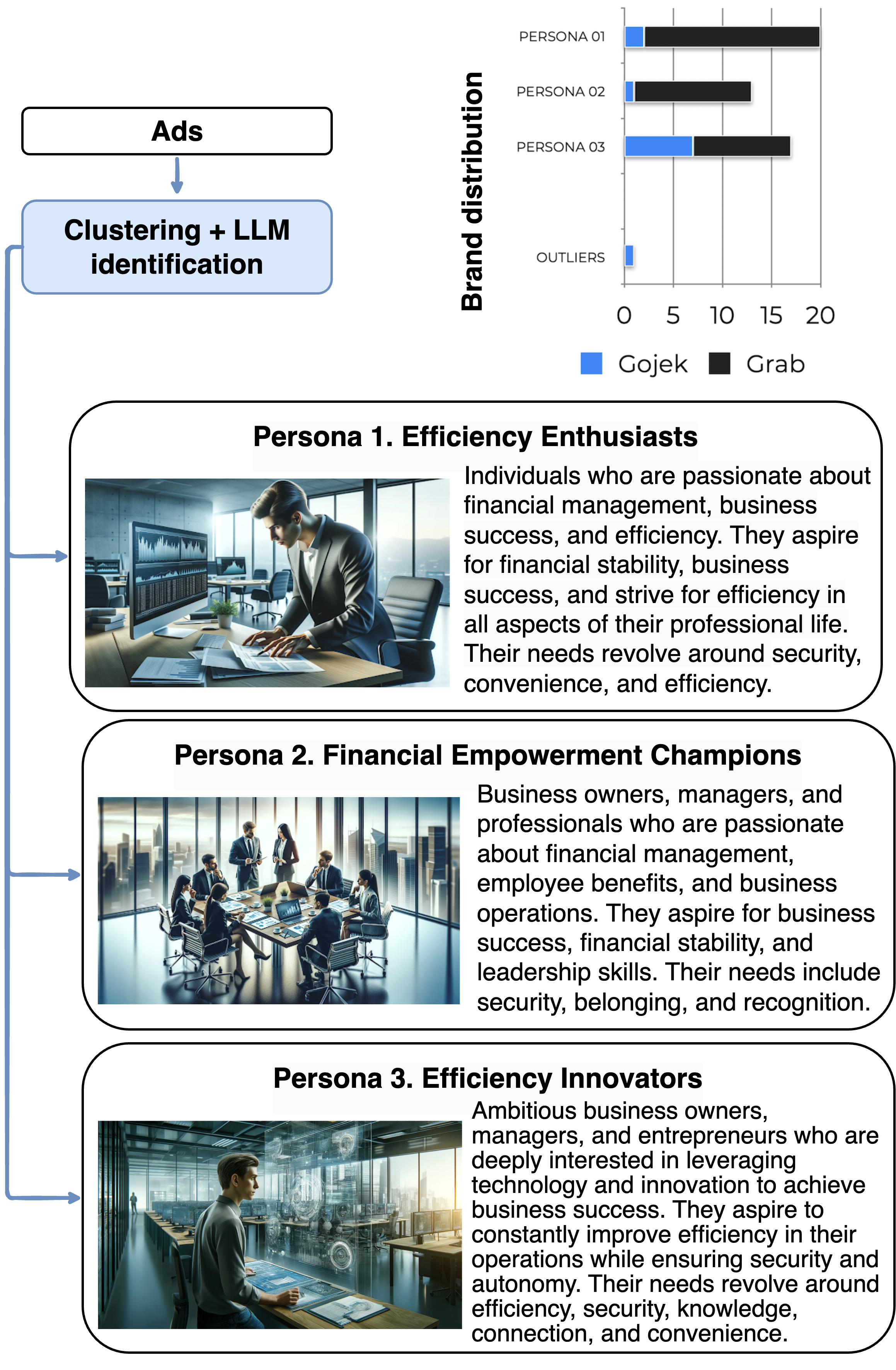} 

\caption{Sample persona analysis results}\label{fig:personas}
\end{figure}

Next, we annotated the clusters, generating names and descriptions by prompting an LLM to do so based on the predominant parameters and content of the posts associated with each cluster. Finally, an image for each persona was generated by the LLM in accordance with the persona's description.
Fig.~\ref{fig:personas} shows the resulting three distinct clusters, each representing a unique customer persona along with their associated characteristics: ``Efficiency Enthusiasts'' (Persona 1), ``Financial Empowerment Champions'' (Persona 2), and ``Efficiency Innovators'' (Persona 3), with content counts of $206$, $144$, and $707$ respectively. Note that some content items did not correspond to any of the three identified personas (i.e., represented outliers) and were thus excluded from the final tally, explaining the disparity with the initial $1{,}120$ ads.

As a result, notable distinctions between the personas have emerged. Both Persona 1 and Persona 3 focus on organizational efficiency, albeit through different approaches: Persona 1 advocates for efficiency through traditional methodologies, while Persona 3 champions it through innovative strategies. In contrast, Persona 2 stands apart by primarily emphasizing the need for financial empowerment of employees.

\subsection{Communication Theme: Challenges}\label{sec:theme}

A persona is an essential but not sufficient component needed to come up with a marketing story and content briefs. In particular, it is important to identify a communication scheme to be combined with the persona's key attributes such as background or character, so that the story would get proper context and would grab attention~\cite{malik2019persona}.
To explore this, 
we have performed X-Means clustering based on the ``Insights'' content pillar, again using ADA embeddings for text~\cite{neelakantan2022text} and running X-Means with $3$ starting centroids and $k_{\max}=50$.

\begin{figure}[!t]\centering
\includegraphics[width=.9\linewidth]{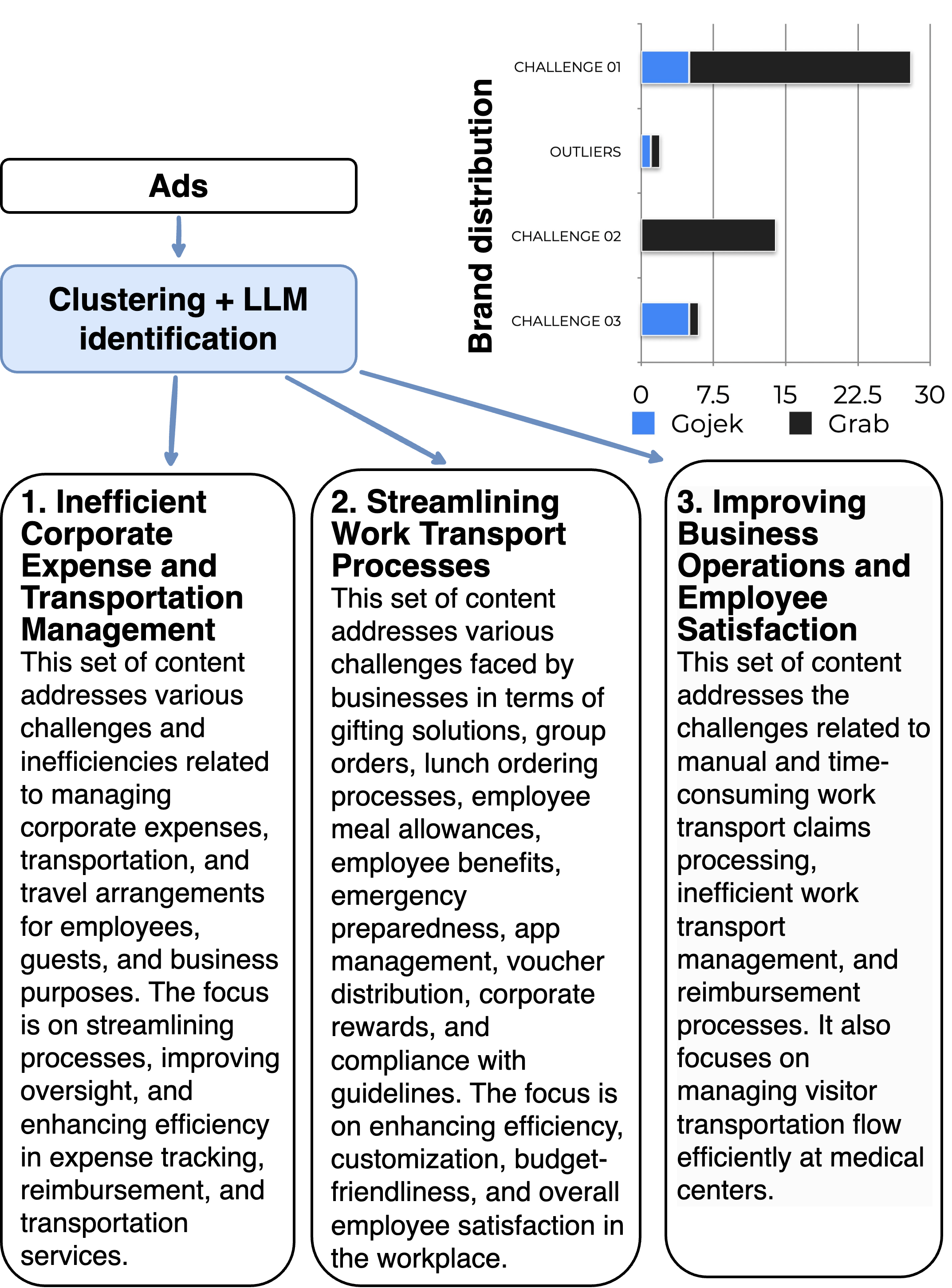} 

\caption{Sample challenges mined by our system}\label{fig:challenges}
\end{figure}

Similarly to persona clustering, we annotated the resulting challenge clusters and generated parameters such as cluster name and cluster description by prompting an LLM to do so based on predominant parameters and content of the posts associated with each cluster. Figure~\ref{fig:challenges} shows sample results: in this case, we have identified three distinct clusters, each representing a unique customer challenge along with their associated characteristics: ``Inefficient Corporate Expense and Transportation Management'' (Challenge 1), ``Improving Business Operations and Employee Satisfaction'' (Challenge 2), and ``Streamlining Work Transport Processes'' (Challenge 3), with content counts of $672$, $94$, and $336$ respectively.

\section{Data-Driven Story Generation}\label{sec:story}

\begin{figure}[!t]
\includegraphics[width=\linewidth]{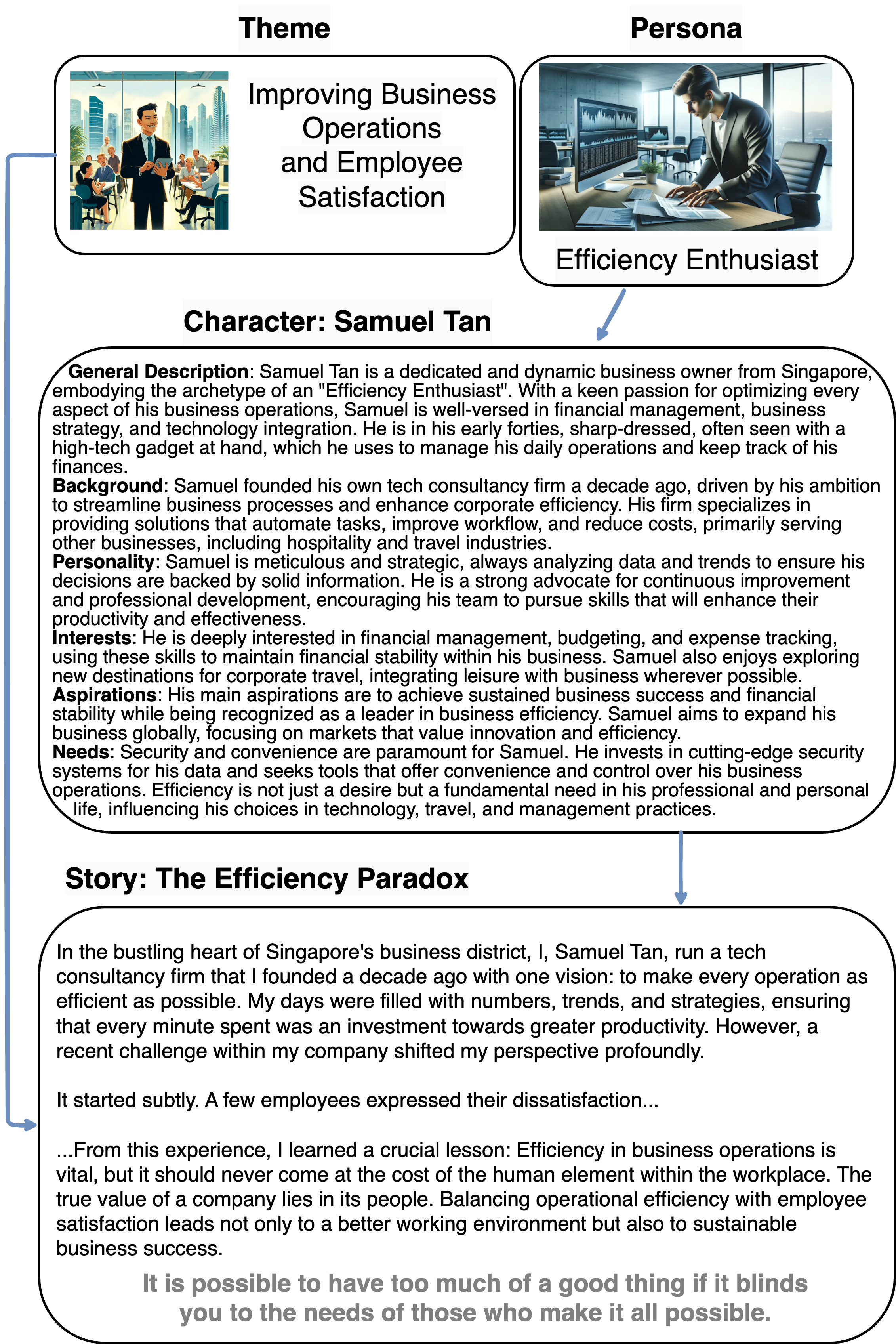} 

\caption{User story generation: pipeline and an example.}\label{fig:story-sample}
\end{figure}

In order to establish a data-driven narrative that would be useful, easy to understand, and attractive to content marketing professionals, it is essential to delineate the content and marketing strategies a brand may deploy, guided by the analysis of the best-performing content from competitive brands. At this point, we have already discussed how the \somon framework extracts personas and challenges based on advertising content prefiltered by a CTR prediction and ranking model. Now it is time to combine them into a unified marketing story.

Consider the scenario where we are part of a marketing team at \gojek, tasked with identifying a suitable communication theme and a target persona for a campaign aimed at market expansion through new audience engagement. It is desirable to opt for a persona and theme that are presently underexploited, thereby expanding our audience reach. As illustrated in Fig.~\ref{fig:personas}, the analysis shows that Persona 1, labeled ``Efficiency Enthusiasts'', is more heavily leveraged by \grab than by \gojek, indicating a potential niche opportunity for \gojek.
At the same time, as shown in Fig.~\ref{fig:challenges}, the communication theme ``Challenge 3: Streamlining Work Transport Processes'' is more frequently addressed by our competitor, \grab, but is underrepresented in \gojek's content. This observation advocates for adopting this theme for persona-centric narrative generation.

Consequently, adhering to the framework outlined in Figure~\ref{fig:story-sample}, we prompt an LLM to amalgamate the characteristics of the selected persona and the chosen communication theme to devise a narrative for a fictitious character, ``Samuel Tan'', a business owner in Singapore. The pipeline for this generation is shown in Figure~\ref{fig:story-sample}: character description is also generated by a custom-prompted LLM, and then the character and theme are combined to create a compelling story. The story is abbreviated in Fig.~\ref{fig:story-sample}, in reality it explains over several paragraphs how Samuel prioritizes business efficiency but often neglects employee satisfaction, leading to increased turnover rates. Recognizing this issue, Samuel opts for \gojek not merely as an efficiency solution but as a means to reengage with employees, enhance their efficiency, and improve overall job satisfaction; the story ends with a conclusion (emphasized in Fig.~\ref{fig:story-sample}) that shows an important insight that Samuel has obtained by engaging with \gojek.
This narrative could serve as a content brief for a creative agency or be utilized as a pitch for a novel marketing campaign, previously unexplored by the marketing team in such depth and with such rapid production capability. In our professional opinion, content briefs produced by \somon are on par with similar human-written narratives and can be used just as effectively.

\section{Case Study: Data-Driven Story Generation}
This section presents a case study evaluating the impact of the proposed story generation approach on marketing workflows across four organizations: NEO360, Blak Labs, Bit by Bit Marketing, and Mothercare. These organizations implemented the system, documented the time required for routine tasks before and after adopting the story generation approach, and reported substantial reductions in time spent on key tasks such as content strategy, campaign planning, competitor monitoring, and audience research.
%
This evaluation was conducted across multiple brands within agency settings and for various product categories at Mothercare.

\begin{table}[!t]
\centering
\caption{Time Savings with Semantic Enrichment Approach}\label{tab:time_savings_icde}
\begin{tabular}{lcc}
\toprule
\textbf{Task} & \textbf{Before (Hours)} & \textbf{After (Hours)} \\ \midrule
\multicolumn{3}{c}{\textbf{NEO360 (Performance Agency)}} \\\midrule
Social Media Proposal Preparation & 3.0 & 0.75 \\ 
Competitor Monitoring & 3.0 & 0.25 \\ 
Audience Research & 5.0 & 0.50 \\ 
Brainstorming Ad Angles & 2.5 & 0.25 \\ \midrule
\multicolumn{3}{c}{\textbf{Blak Labs (Creative Agency)}} \\\midrule
Campaign Planning & 5.0 & 2.0 \\
Content Planning & 10.0 & 3.0 \\
Brand Analysis & 10.0 & 1.0 \\ \midrule
\multicolumn{3}{c}{\textbf{Bit by Bit Marketing (Branding Agency)}} \\\midrule
Content Strategy & 5.0 & 1.0 \\ \midrule
\multicolumn{3}{c}{\textbf{Mothercare (Consumer Brand)}} \\\midrule
Competitor Analysis & 5.0 & 2.0 \\ 
Ad Compilation & 5.0 & 1.0 \\ 
Customer Segmentation & 5.0 & 2.0 \\ \bottomrule
\end{tabular}
\end{table}


\subsubsection{NEO360: Streamlining Performance Marketing}
NEO360\footnote{\url{https://neo360.digital/}}, a performance agency, implemented the semantic enrichment approach to enhance workflows for tasks such as social media preparation, competitor monitoring, audience research, and brainstorming ad angles. As summarized in Table~\ref{tab:time_savings_icde}, the system reduced task durations by 2x to 3x;
e.g., the time required for social media proposal preparation was reduced from 3 hours to 45 minutes, while competitor monitoring was shortened from 3 hours to just 15 minutes, 
showing radical acceleration of research-intensive tasks.

\subsubsection{Blak Labs: Optimizing Creative Workflows}
Blak Labs\footnote{\url{https://blaklabs.com/}}, a creative agency, leveraged the system to streamline campaign planning, content planning, and brand analysis. The results, shown in Table~\ref{tab:time_savings_icde}, demonstrate time reductions of 2x to 3x.
Notable gains include reducing the time for brand analysis from 10 hours to 1 hour and content planning from 10 hours to 3 hours, proving that the platform can 
handle resource-intensive creative processes.

\subsubsection{Bit by Bit Marketing: Enhancing Branding Strategies}
Bit by Bit Marketing\footnote{\url{https://bitbybitmarket.com/}}, a branding agency, used the semantic enrichment approach to optimize its content strategy workflow. Table~\ref{tab:time_savings_icde} shows that the agency achieved a 5x reduction in time spent on this task, from 5 hours to 1 hour; in this case, the system was able
to streamline strategic processes, enabling agencies to focus on higher-level planning and execution.

\subsubsection{Mothercare: Improving Consumer Brand Efficiency}
Mothercare\footnote{\url{https://www.mothercare.com.sg}}, a consumer brand, applied the system to optimize competitor analysis, ad compilation, and customer segmentation workflows. Table~\ref{tab:time_savings_icde} highlights the 3x to 5x improvements achieved; e.g.,
ad compilation time was reduced from 5 hours to 1h, and customer segmentation tasks were completed in less than half the time. These results validate our approach’s versatility in enhancing productivity across diverse domains.

Results of this case study highlight the transformative impact of data-driven story generation in marketing workflows. 
Our LLM-based system not only reduces task durations but also provides actionable insights for optimizing operations. The proposed platform can 
address real-world challenges and enable marketers to achieve huge productivity gains while maintaining high-quality outcomes. This evidence strongly supports the adoption of semantic enrichment and data-driven content optimization in marketing practices.

\section{Content scoring}\label{sec:ctr}

\subsection{Problem setting and dataset}
To make actionable recommendations, one has to enable marketers to evaluate data and select which content to incorporate into analytics, ideally based on anticipated performance. This leads to the need for content scoring, so in this section we add to \somon a system for the scoring of advertising content, developed in our previous work~\cite{10.1145/3581783.3612817}, and evaluate it against state of the art LLMs to prove its continued relevance.


\begin{table}[!t]\centering
\caption{Experimental results for \somon.}\label{tbl:soda}
\begin{tabular}{lccccc}
\toprule
  Brand        & nDCG@5 & nDCG@10 & Recall@3 & Recall@5 \\ \midrule
A  & 0.498  & 0.796 &    0.333 & 0.4\\ 
 B  & 0.541  & 0.821 &  0.333 & 0.4 \\ 
 C  & 0.588  & 0.829 &  0.6 & 0.6\\ \bottomrule
\end{tabular}
\end{table}

\begin{table}[!t]\centering
\caption{Performance evaluation against LLM baselines; ``gd'' represents a grounded model.}\label{tbl:llms}
\begin{tabular}{lcccc}
\toprule
              & nDCG@5         & nDCG@10        & Recall@3     & Recall@5 \\ \midrule
GPT4-gd       & 0.591          & 0.818          & 0.333        & 0.6      \\
GPT4          & 0.465          & 0.755          & 0            & 0.4      \\
GPT4o-gd      & \textbf{0.599} & 0.827          & 0.333        & 0.6      \\
GPT4o         & 0.511          & 0.750          & 0.333        & 0.4      \\
Gemini-gd     & 0.578          & 0.817          & 0.333        & 0.6      \\
Gemini        & 0.564          & 0.769          & 0            & 0.6      \\
\somon & 0.588          & \textbf{0.829} & \textbf{0.6} & 0.6      \\ \bottomrule
\end{tabular}
\end{table}

To assess the effectiveness of our recommender system, we conducted a validation study using real ad creatives and their performance data. We collected real-world ad creatives and their performance data from three brands on \emph{Facebook}, each with distinct marketing objectives: sales, conversion, and traffic, spanning from October 1, 2023 to December 31, 2023. The dataset comprised text and images of their ads alongside their performance measured as CTR values. We chose three (anonymized) brands: Brand A (16 ads in the fitness industry), Brand B (29 ads in the education industry), and Brand C (16 ads in the automobile industry).

The ads were assessed not just according to the brand affiliation, but also according to the Ad objective selected on the Meta Ads platform, which is a crucial component of evaluation in our scenario due to the fundamental differences in click-through rates across various industries and marketing goals that Ads achieve with different Ad objective settings~\cite{10.1145/3581783.3612817}. For instance, advertisements optimized for clicks may result in lower costs per click and higher click-through rates, yet they may not effectively convert customers. Conversely, ads aimed at achieving better conversion rates might not exhibit as high click-through rates as those optimized merely for clicks.

\subsection{Content scoring models}
We use our previously developed content scoring approach~\cite{10.1145/3581783.3612817}; it was originally designed to predict CTR by classifying ads into CTR performance categories but here we operate in a recommendation setting, i.e., pose content scoring as a ranking problem because
\begin{inparaenum}[(i)]
    \item marketers do not need precise CTR values to choose the best content but rather a comparative assessment,
    \item CTR values undergo significant distribution shift across platforms, time, and other external factors, but the ranking of ads should be more stable, and
    \item the LLM performance in qualitative ranking should be significantly better than in numerical predictions.
\end{inparaenum}

Thus, we take the SODA model from~\cite{10.1145/3581783.3612817} that predicts a distribution over ``High CTR'', ``Average CTR'', and ``Low CTR'' labels and add a linear layer that translates the probabilities into a single score via a trainable linear combination:
$$\mathrm{Score} =  \alpha \cdot p(\mathrm{HighCTR}) + \beta\cdot (1 - p(\mathrm{HighCTR})),$$
converting a classification model into a ranking model.





On the side of LLMs, we used GPT-4~\cite{openai2023gpt4}, its latest multimodal variant GPT-4o, and Gemini 1.5 Pro~\cite{geminiteam2024gemini} for comparison with \somon.
%
A significant challenge in utilizing LLMs for this task is their tendency to hallucinate, which can result in generating inaccurate or nonsensical information that compromises the reliability of their output. To mitigate this issue and ground the LLM-based approach to ensure the predicted data was trustworthy, we implemented several measures to reduce randomness:
\begin{inparaenum}[(i)]
    \item \emph{temperature adjustment}, setting the temperature parameter of the LLMs to 0.1, which is known to improve stability by reducing the variability in the model's outputs;
    \item \emph{grounding}, i.e., including another three sample ads from the brand within the prompt, each with a specified CTR rank: one with the best performance, one with average performance, and one with the worst performance;
    \item \emph{ensembling}, 
    running the ranking process with LLMs five times in the same settings and ranking the ads by the (lowest to highest) sums of their ranks from all runs.
\end{inparaenum}

\subsection{Results}

For evaluation, we use standard ranking-based metrics, namely
Recall@3 and Recall@5.
Recall is very important in practice, especially in resource-constrained domains such as online advertising campaigns, where 
top $k$ recommendations should encompass a substantial portion of the truly relevant items, so that marketers can
focus their efforts on the most promising opportunities.
We have also computed normalized discounted cumulative gain (nDCG) at cutoffs 5 and 10, another commonly used way to assess ranking quality.
%
%
Table~\ref{tbl:soda} shows a case study 
across three major industries. In the fitness sector, \somon achieved moderate effectiveness with nDCG@5 of 0.498 and nDCG@10 of 0.796, alongside Recall@3 and Recall@5 scores of 0.333 and 0.4, respectively. Brand B, representing education, showed improved performance, with nDCG@5 of 0.541 and nDCG@10 of 0.821, but recall stayed the same.
In the automobile industry results improved, achieving nDCG@5 of 0.588 and nDCG@10 of 0.829, with recall@3 and recall@5 both scoring 0.6. These findings show that content scoring quality varies across industries and highlight the importance of industry-specific optimization to enhance recommendation accuracy.

Then we evaluated \somon against LLM baselines.
In campaigns focused on driving traffic, the primary goal is to maximize click-through rates (CTR), which serve as a measure of user engagement and interaction with the advertisement, which is particularly pertinent given the distinctive challenges posed by traffic-driven campaigns as opposed to conversion-oriented strategies. Thus, we evaluate the performance with the data from Brand C, a traffic-oriented campaign. Table~\ref{tbl:llms} presents the results from a case study focused on Brand C, comparing the performance of \somon against modern LLMs.
It shows a significant performance boost achieved by knowledge grounding across various models and metrics; e.g., nDCG@5 for GPT-4 improved from 0.465 without grounding to 0.591 with grounding, while its Recall@5 increased from 0.4 to 0.6. This also holds for 
GPT-4o and Gemini1.5 Pro, showing that grounding is critical in enhancing the relevance and quality of recommendations.
Ungrounded LLMs exhibit markedly lower performance, particularly in the Recall@3 metric: both Gemini and GPT-4 failed to place any high-performance items within their top $3$ results at all.

As a result, the prediction model outperforms LLMs in recommending relevant items, achieving the highest nDCG@10 of 0.829 and recall@3 of 0.6 among compared models. \somon is better able to balance ranking accuracy and item recall, delivering highly relevant and stable (especially compared to individual LLM results) recommendations to users. Notably, these outcomes are achieved without any fine-tuning (grounding) using data from the actual industry-specific distributions: we used a pretrained model. This offers significant potential for further improvement of results should such fine-tuning be implemented, a direction we intend to explore in our future work. Improvements in LLMs' Recall@k due to grounding also hint at the critical role of domain knowledge and context in enhancing the relevance and quality of recommendations. Ungrounded LLMs, relying solely on their pre-trained language models, struggle to capture the nuances and specifics of the recommendation task, resulting in sub-optimal item retrieval. By incorporating domain-specific knowledge and grounding the models in the relevant context, the recommendations become more tailored and relevant.

We note that our proposed \somon framework outperforms the baselines in both Recall@3 and Recall@5, achieving a Recall@3 score of 0.6 which is higher than any LLM's result. This supports our conclusion that the online marketing industry still cannot fully rely on LLMs for content scoring and explains our decision to include a CTR prediction and ranking model into the mostly LLM-based \somon framework.

\section{Quantitative Evaluation and Case Study}\label{sec:case_study}

\begin{figure*}[!t]\centering
\includegraphics[width=.9\linewidth]{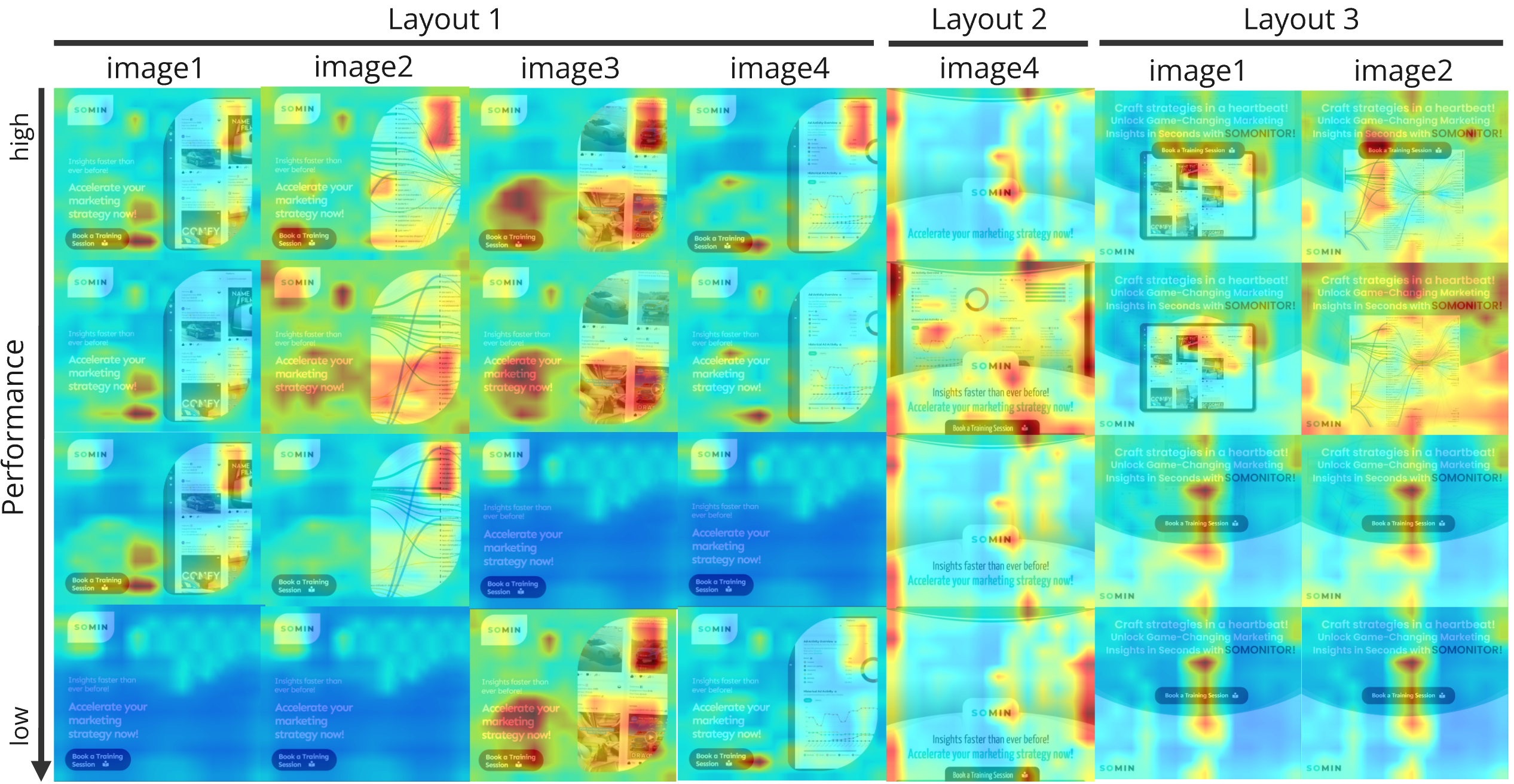}
\caption{Visualization of object removal guided by attention heatmaps.}
\label{fig:heatmap}
\end{figure*}

\begin{figure*}[!t]
\includegraphics[width=\linewidth]{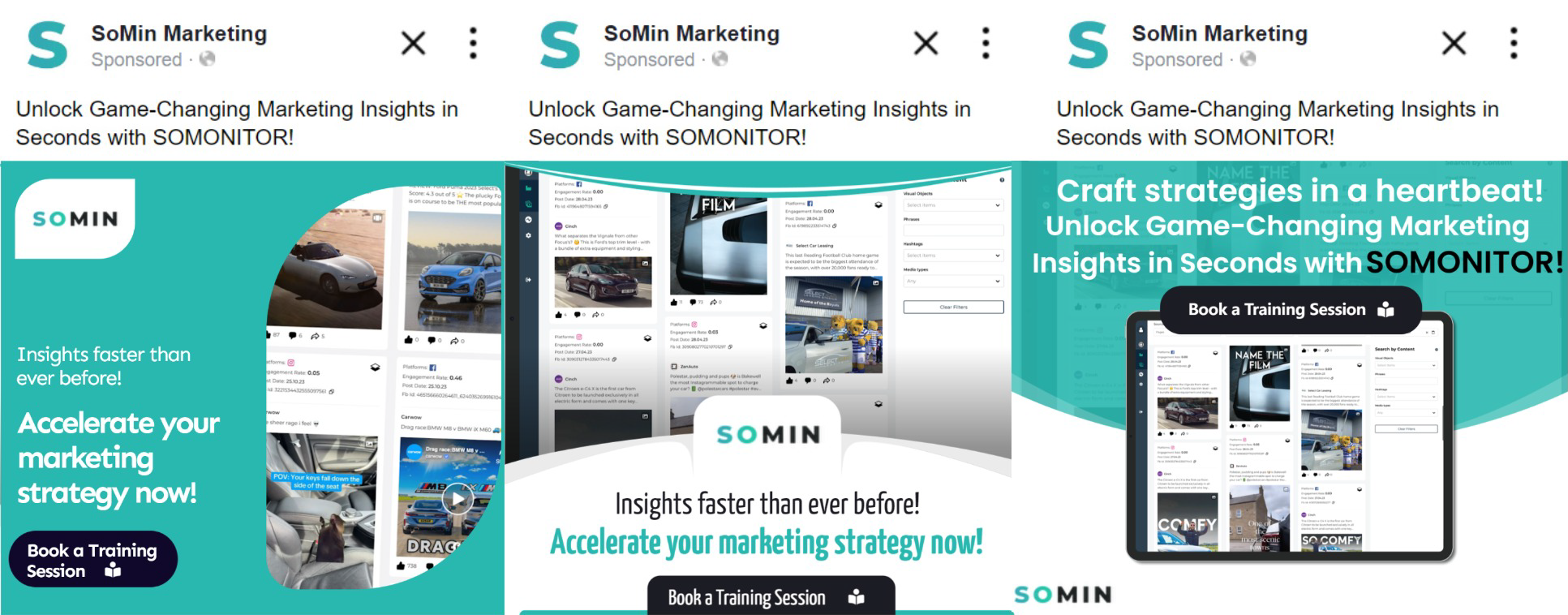}
\caption{Example of an ad variant deployed on the Meta platform.}
\label{fig:ad_variant}
\end{figure*}

We have shown that a small neural network can outperform one-shot fine-tuned LLMs in predicting content performance prior to the official release of advertisements. However, for this solution to be practical for marketers it must not only identify what works better but also provide actionable guidance on how to improve the content. We have noted before~\cite{10.1145/3581783.3612817} that attention layers of the content scoring network can be visualized to assist marketers in refining their content iteratively, but have not tested it in real world conditions until now. In this section, we evaluate this approach with the \emph{Meta} advertising engine, aiming
(1) to assess whether regions highlighted by attention heatmaps correlate with higher user engagement and (2) to determine whether removing key objects identified by heatmaps adversely affects campaign performance.

\subsection{Experimental Setup and Methodology}

We used a diverse dataset of advertising creatives with
multiple visual elements and
processed them through our predictive framework, which generated attention heatmaps to identify areas likely to maximize user engagement.

\emph{Selection of Creative Assets:} only creatives with high-confidence predictions (above the model’s average confidence threshold) were included in the experiment. This ensured that selected assets featured impactful elements, allowing for a focused assessment of the heatmap’s predictive capabilities.

We used the following experimental workflow:
\begin{enumerate}
    \item \emph{heatmap generation:} each creative was processed with \sowide, to generate heatmaps (Fig.~\ref{fig:heatmap}), highlighting elements expected to drive user engagement;
    \item \emph{iterative object removal:} based on the heatmaps, critical objects were sequentially removed, and updated heatmaps were generated after each iteration; this process produced three modified versions of each original creative;
    \item \emph{campaign deployment:} original and modified creatives were deployed in a real-world campaign on the \emph{Meta} platform with identical settings (\textit{Objective}: Landing Page View; \textit{Placements}: All; \textit{Budget Control}: automatic; \textit{Ad Copy}: "Unlock Game-Changing Marketing Insights in Seconds with SOMONITOR"; \textit{Duration}: 1 day);
    \item \emph{performance metrics:} key performance indicators such as click-through rate (CTR), click-to-landing page view (CTR-LPV), and landing page views (LPV) were monitored for all creatives, with a total of 65{,}107 impressions, 2{,}243 clicks, and 1{,}162 landing page loads recorded;
    \item \emph{performance analysis:} 
    to quantify the impact of object removal, we counted 
    the proportion of cases where performance declined after object removal.
\end{enumerate}

\subsection{Results and Observations}

\begin{table}[!t]
\centering\setlength{\tabcolsep}{8pt}
\caption{Evaluating object removal in a real-world campaign}
\begin{tabular}{lcccc}
\toprule
\textbf{Layout} & \textbf{LPV} & \textbf{CTR-LPV} & \textbf{CTR} & \textbf{F1 Score} \\ \midrule
Layout 1 & 0.769 & 0.769 & 0.692 & 0.750 \\
Layout 2 & 0.751 & 0.751 & 0.250 & 0.500 \\
Layout 3 & 0.570 & 0.857 & 0.857 & 0.860 \\ \midrule
\textbf{Overall} & 0.708 & 0.792 & 0.667 & 0.740 \\\bottomrule
\end{tabular}
\label{tab:evaluation}
\end{table}

Experimental results summarized in Table~\ref{tab:evaluation} show the efficacy of heatmaps for guiding ad optimization:
\begin{itemize}
    \item \emph{Layout 1:} Moderate performance with well-aligned LPV and CTR-LPV values (0.769), indicating consistent user engagement. The F1 score (0.750) suggests accurate predictions of ad performance.
    \item \emph{Layout 2:} Significantly lower CTR (0.250) shows problems in attracting initial clicks despite robust post-click engagement (CTR-LPV 0.751), reflecting the model's limitations in detecting less impactful visual cues.
    \item \emph{Layout 3:} The highest CTR-LPV (0.857) and F1 score (0.860) demonstrate strong alignment between predicted and actual user engagement. However, lower LPV (0.570) indicates potential gaps in predicting post-click retention.
\end{itemize}

\subsection{Impact of Object Removal}

\begin{figure*}[!t]\centering
\includegraphics[width=.9\linewidth]{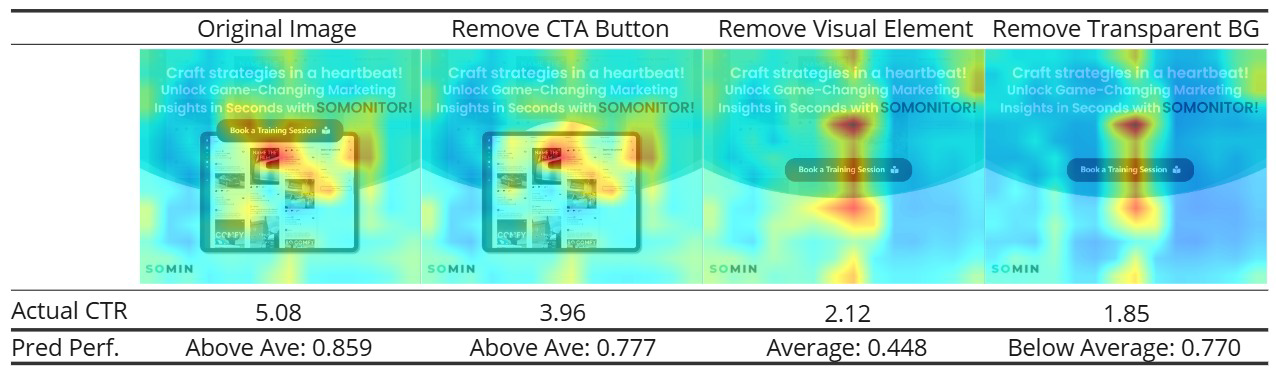}
\caption{Performance comparison of creatives after object removal.}
\label{fig:comparison_ads}
\end{figure*}

Systematic removal of key visual elements led to the following results:
\begin{inparaenum}[(i)]
    \item removing the \emph{CTA Button} caused a 22\% drop in CTR, showing its importance in driving engagement.
    \item removing \emph{transparent background} led to the most severe performance decline, with a 64\% drop in CTR, highlighting its influence in guiding user attention;
    \item removing the \emph{primary visual element} resulted in a 58\% CTR drop, demonstrating its significance in capturing initial user attention.
\end{inparaenum}
The model’s predictions are closely aligned with observed performance.

\section{Gen-AI Performance Marketing Consultant}\label{sec:optimization}

We have been exploring strategies that enable marketers to enhance their content efforts, as well as prevalent trends in content marketing. 
However, major companies like \emph{Procter \& Gamble} or \emph{McDonald's} often undergo rigorous reviews before altering creative materials or ad copy. This can make proposed improvements to advertising content impractical, as the targeted campaign may end before suggestions are implemented. When creative changes are unfeasible, real-time campaign optimization becomes essential. Even for campaigns with high-quality creatives, this approach ensures precise alignment of creatives with target audiences and helps identify performance bottlenecks through detailed data analysis.
In this section, we discuss how LLM-based analysis can help evaluate key performance metrics and guide strategic decisions in campaign optimization, connecting the micro-level analysis of ad creatives with the macro-level approach of the LLM helping steer the course of the entire advertising campaign. 

\subsection{Metrics for Measuring Campaign Performance}\label{sec:metrics}

We expand the focus from ad creative analysis to a holistic view of campaign performance using several key metrics:

\begin{itemize}
    \item \emph{Reach}: total unique individuals exposed to the campaign, indicating its audience breadth and potential impact on brand visibility and awareness.
    \item \emph{Frequency}: average number of times a user sees the ad; tracking it is needed to avoid audience fatigue~\cite{abrams2007personalized};
    \item \emph{Results}: total targeted conversions (e.g., leads or purchases), the primary measure of campaign success;
    \item \emph{Cost per Result}: average cost per conversion, reflecting campaign efficiency and guiding budget optimization;
    \item \emph{Amount Spent}: total expenditure, needed for budget tracking and ensuring alignment with campaign goals;
    \item \emph{CPM} (Cost per 1{,}000 Impressions):
    cost of reaching 1{,}000 users informs about market competition~\cite{aggarwal2009general};
    \item \emph{CTR} (Click-Through Rate), the prediction target for our \sowide model (Section~\ref{sec:ctr}); it assesses ad engagement and content effectiveness;
    \item \emph{CR for Link Clicks to LP Views}: fraction of users who click the ad and view the landing page, indicating landing page effectiveness;
    \item \emph{CR for Link Clicks to Results}: fraction of users clicking the ad who achieve the targeted conversion, directly reflecting campaign goal attainment.
\end{itemize}

These metrics collectively provide a comprehensive view of campaign performance, encompassing audience engagement, conversion rates, budget efficiency, and ad quality, enabling informed optimization decisions.

\subsection{Data Acquisition and Preprocessing}\label{sec:trendmeta}

\begin{figure}[!t]
\includegraphics[width=\linewidth]{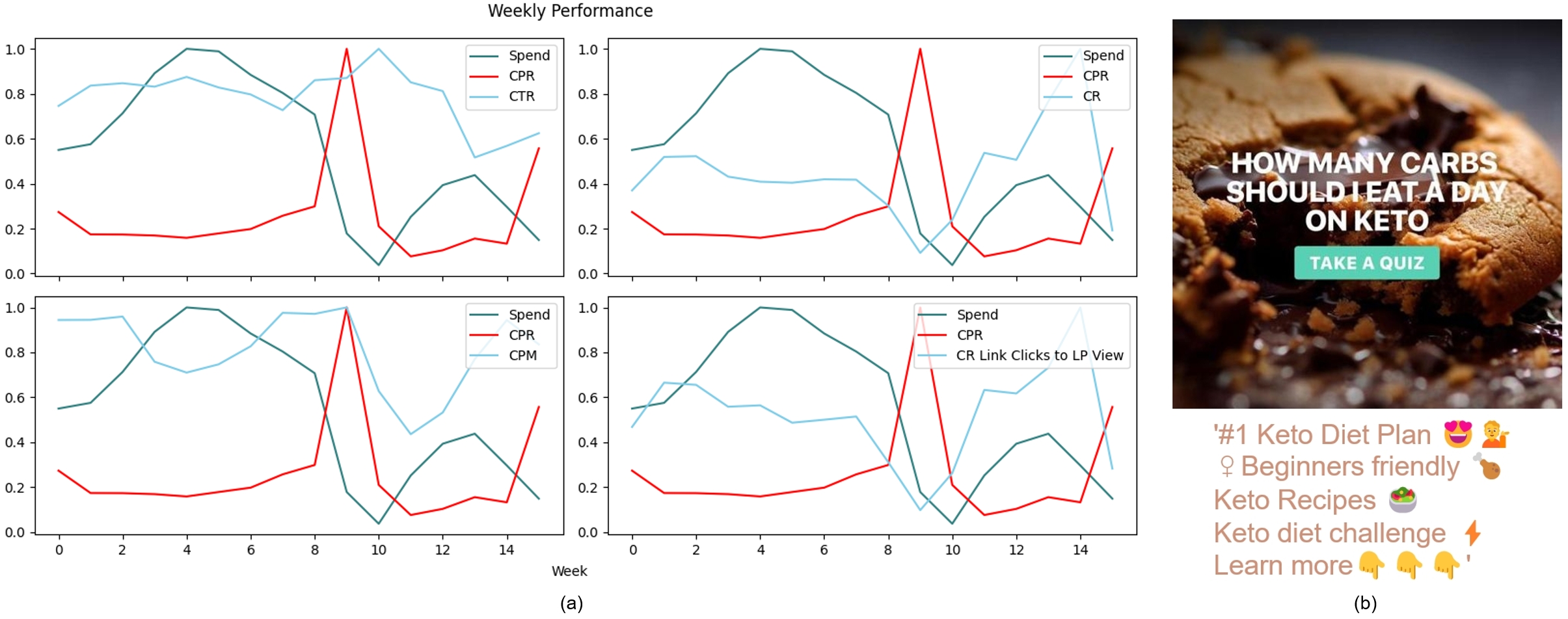} 
\caption{Data for Section~\ref{sec:trendmeta}: (a) weekly performance; (b) a sample creative.}\label{fig:trend}
\end{figure}

Metrics data for the \emph{Meta} campaign were collected over a three-month period (October 1, 2023 – January 23, 2024) at three levels of granularity:
\begin{inparaenum}[(i)]
    \item \emph{weekly}, which provides a consolidated view by smoothing short-term fluctuations 
    (Fig.~\ref{fig:trend});
    \item \emph{daily}, which offers detailed insights into day-to-day variations essential for detecting anomalies or immediate external influences, and at the
    \item \emph{creative level}, which evaluates the contribution of individual creatives to campaign performance.
\end{inparaenum}
%
To adapt data for LLM analysis, it was preprocessed as follows:
\begin{inparaenum}[(i)]
    \item \emph{creative data} represented as tabulated insights with images resized to $512 \times 512$ pixels and encoded in base64;
    \item \emph{numerical metrics}, as percentage change rates between consecutive data points to focus on trends and fluctuations rather than raw values.
\end{inparaenum}
This preprocessing ensures that metrics, diverse in granularity and spanning over 115 days, are strategically well-structured for comprehensive campaign analysis.

\subsection{Prompt Engineering}

Automated campaign performance analysis requires a well-crafted prompt in order to guide the LLM through a comprehensive evaluation of key metrics and their interdependencies. Our prompt format follows the convention of Knowledge + Role + Task + Data and provides the necessary context and information for the LLM's analysis.

\begin{table*}[!t]\small\setlength{\tabcolsep}{6pt}\centering
\caption{Sample ad campaign analysis results by GPT-4}\label{tbl:gptresults1}
\begin{tabular}{p{.47\linewidth}|p{.47\linewidth}}\toprule
\begin{enumerate}[wide, labelwidth=!, labelindent=0pt]
\item \emph{Key Metrics Change}: 
{Both CPR (Cost Per Result) and SPEND did change over the course of the campaign.}
\item \emph{Key Metrics Trend}:
\begin{itemize}
    \item {CPR started at \$57.01, decreased to \$33.08 at its lowest point, spiked to \$208.28, and then decreased again to \$27.68 before rising to \$115.98. The trend indicates a fluctuation in cost efficiency, with an overall decrease towards the middle of the campaign followed by a substantial increase towards the end.}
    \item {SPEND varied significantly, with an initial value of \$2565.39, rising to \$4664.91, then dropping to \$175.18, and finally increasing to \$695.91. The campaign budget was dynamically adjusted, leading to these variations.}
\end{itemize}

\item \emph{Secondary Metrics Trend}:

\begin{itemize}
   \item {CTR (Click-Through Rate) fluctuated, starting at 2.68\%, peaking at 3.59\%, and ending at 2.24\%. Changes in CTR indicate varying levels of user engagement with the ads.}
   \item {CPM (Cost per 1,000 Impressions) decreased to \$9.98, spiked to \$14.06, and then dropped to \$6.13 before rising again to \$11.75. This suggests that ad performance and competition for audience attention were inconsistent.}
   \item {CR (Conversion Rate) and CR Link Clicks to LP View showed fluctuations, with a general downward trend towards the end of the campaign. This could indicate a decrease in ad relevance or landing page effectiveness.}
\end{itemize}
\item \emph{Relation between Key and Secondary Metrics}: 
{The increase in CTR typically suggests more effective ad content, which should lower CPR by generating more leads or purchases for the same or less spend. However, the spike in CPM indicates a higher cost to reach the audience, which can increase CPR if the rise in ad costs is not offset by a proportionate increase in purchases. The downward trend in conversion rates, both overall and from link clicks to landing page views, directly contributes to a higher CPR, as fewer users who click the ad proceed to purchase.}
\item \emph{Impact of Secondary Metrics on Primary Metrics}: 
{CTR directly influences CPR; a higher CTR means more clicks for the budget, increasing the opportunity for conversions and potentially lowering the CPR if the conversion rate is stable or improves. Conversely, an increase in CPM raises the cost to gain impressions, potentially increasing the CPR if not balanced by a rise in conversion rates. The conversion rate from link clicks to landing page views is pivotal; a decrease suggests that users are less compelled by the landing page to make a purchase, thus increasing CPR.}
\end{enumerate}
&
{The correlation is clear: while a higher CTR and improved landing page conversion rates contribute to a lower CPR, increases in CPM or declines in conversion rates can negate these benefits, leading to a higher CPR. The data shows fluctuations in CTR and CPM, alongside a general decrease in conversion rates, explaining the observed changes in CPR.}
\begin{enumerate}[wide, labelwidth=!, labelindent=0pt]
\setcounter{enumi}{5}
\item \emph{Hypothesis for Changes in Primary Metrics}:
\begin{itemize}
    \item {The initial decrease in CPR can be attributed to the campaign's optimization and improvements in ad targeting, leading to a more engaged audience and more efficient spending.}
    \item {The spike in CPR could be due to market dynamics, such as increased competition, leading to higher costs for reaching the audience.}
    \item {The subsequent decrease in CPR and then increase suggests that the audience may have experienced ad fatigue, resulting in lower conversion rates and higher costs per purchase.}
\end{itemize}
\item \emph{Analysis of Textual and Visual Data of Ads}:
\begin{itemize}
   \item {The ad's textual content is clear and direct, with a focus on the Keto diet, which is relevant to the target audience. However, the visual content, while appealing, does not directly link to the desired action of taking a quiz or making a purchase.}
   \item {To improve the key metrics, I recommend testing variations of the ad copy that directly address the benefits of the Keto diet and include a strong call-to-action. Additionally, using visuals that better represent the quiz or the benefits of following the Keto diet may improve conversion rates.}
\end{itemize}

\item \emph{Recommendation for Next Steps}:
\begin{itemize}
   \item {Monitor the CTR and conversion rates closely and test different ad creatives to find the most engaging content for the audience.}
   \item {Analyze the times when CPM spikes to determine if there are patterns related to increased competition and adjust bidding strategies accordingly.}
   \item {Review the landing page experience to ensure it is optimized for conversions, with fast load times and clear calls-to-action.}
   \item \textit{Consider retargeting campaigns to engage users who have previously interacted with the ads but have not converted, using tailored messaging based on their interaction stage.}
   \item {Allocate budget dynamically, but with caution during periods of high CPM to avoid inefficiency, focusing more on periods with lower CPM and higher engagement.}
\end{itemize}
\end{enumerate}
\\\bottomrule
\end{tabular}
\end{table*}

\subsubsection*{Knowledge of Campaign Metrics} 

Evaluation metrics span different facets of an ad campaign, from audience engagement to cost efficiency; they are explained in detail in the prompt and include Reach, Frequency, Results, Cost per Result, Amount Spent, CPM, CTR, CR Link Clicks to LP View, and CR Link Clicks to Result (see Section~\ref{sec:metrics}). Each metric serves a distinct purpose in evaluating campaign performance and offers valuable insights into audience interaction, cost-effectiveness, and conversion rates.

\subsubsection*{Role of the LLM} 

The prompt also defines the specific role that the LLM is supposed to play, namely operate as a surrogate performance marketing manager. The LLM is expected to embody confidence in its analyses, refraining from speculative language such as ``could be'' or ``might be''. In this simulated role, the LLM is entrusted with the responsibility of scrutinizing performance metrics and providing well-founded explanations for observed trends. This confidence aligns with the typical expertise and authority of a seasoned performance marketing professional, which improves the resulting analysis.

\subsubsection*{Task Definition for the LLM}

The task assigned to the LLM is explicitly outlined in the prompt. The primary objective is to achieve a lower CPR, and the manager dynamically adjusts the budget to optimize this metric. The task involves not only tracking changes in key metrics such as CPR and Spend but also understanding the trends in secondary metrics such as CTR, CPM, CR, and CR Link Clicks to LP View.

\subsubsection*{Key Questions for LLM Analysis}

To guide the LLM through a systematic analysis, we present in the prompt a set of key questions inspired by the routine thought process of an ad optimization manager. These questions include the following.

\begin{enumerate}
 \item Did the key metrics (CPR, SPEND) change?
 \item {If they changed, summarize the trend of the key metrics.}
 \item {What was the trend of secondary metrics (CTR, CPM, CR, CR Link Clicks to LP View) during the same time?}
 \item {Are the changes in key and secondary metrics related?}
 \item {Which secondary metrics influence the key metrics?}
 \item {Provide hypotheses to explain the changes in key metrics.}
 \item {Analyze textual and visual data of ads to explain key metric changes and recommend improvements.}
 \item {Recommend the next steps for campaign performance improvement based on the metrics.}
\end{enumerate}

\subsubsection*{Tabular Data Integration}
The prompt is enriched with tabular data that contains both numerical values of metrics and detailed creative features of the ads. This allows the LLM to analyze both quantitative and qualitative aspects, grounding its understanding of the campaign performance and helping generate actionable recommendations for optimization.

\subsubsection*{Visual Data Integration}
In addition to text and tabular data, we included the ads' visuals to enrich contextual information within the prompt. Thus,
a multimodal LLM such as GPT-4 (that we used in the experiments) 
can not only comprehend textual nuances but also mine insights from the visual contents of the campaign, again leading to a more holistic understanding of the relations between creative content and performance.

\subsection{Analysis of the Results}

Here, we present the LLM-based analysis for performance metrics evaluation in the specified marketing campaign. The examination primarily focuses on key metrics, such as CPR and Amount Spent, alongside with secondary metrics like CTR, CPM, CR, and CR Link Clicks to LP View. The goal is to uncover insights into the dynamics of the campaign and provide actionable recommendations for improvement.
LLM results are shown in Table~\ref{tbl:gptresults1}.
The LLM's outputs demonstrate its proficiency in annotating data and accurately detailing variations in key metrics, as seen in points 1, 2, and 3. This capability is important for understanding recent campaign performances and for summarizing critical data for marketing managers in preparation for reporting meetings. Moreover, the model is quite adept in identifying the root causes of increasing the cost per result metric. It points out the decline in conversion rates and the rise in CPM in points 4 and 5 as contributing factors.
In point 6, the LLM goes further, recognizing potential issues such as ad fatigue, ad targeting optimization, and market seasonality. This is especially important because these aspects were not explicitly input in the prompt or data, indicating the model's ability to extract new insights. In a manner similar to the analysis in Section~\ref{sec:ctr}, point 7 discusses potential enhancements in visual content and explores the reasons for its current inefficiency.
However, point 8 shows a limitation: suggestions provided there are quite generic and may not fully align with the specific needs of the keto diet campaign. These recommendations should be customized to better suit the unique aspects of the campaign.
Despite these limitations, the analysis provided by the LLM in this example proves to be a valuable asset for digital marketing professionals, offering quick evaluations and identifying potential areas for optimization across various campaigns. Initial recommendations provided by our platform are especially helpful for early-career marketers and serve well for educational purposes. However, more advanced marketing professionals may require more specific and detailed insights, which we plan to explore in future research.

\section{Conclusion}\label{sec:concl}

In this work, we have introduced an innovative explainable AI framework  \somon that aims to combine human intuition with the efficiency of modern AI to significantly enhance the capabilities of marketers across all stages of the marketing funnel, from strategic planning to content creation and campaign execution. The challenges traditionally faced by marketers, especially the need to manually sift through immense quantities of transient online content, are addressed by \somon's ability to automate and refine the process of competitor analysis, content research, and strategic branding, which has been proven by a case study in the real-world settings.
\somon ranks advertising content by its potential to achieve higher click-through rates via a specially trained model and then employs LLMs to distill core content pillars, which lead to communication themes and targeted customer personas, providing a structured basis for content and campaign planning.
By integrating these insights with brand marketing data, \somon not only constructs compelling narratives for new customer personas but also generates detailed content briefs in the form of customer persona stories, empowering marketing teams to produce content and execute campaigns in a guided manner, which has been proven to help in prioritizing different content elements and enhance real-world advertising performnace. We believe that \somon can transform digital marketing operations, enabling marketers to quickly navigate through large datasets and derive actionable insights that boost campaign effectiveness and enhance job satisfaction, redefining traditional marketing approaches~\cite{farseev2024transparency}.

\section{Acknowledgement}\label{sec:concl}
This work was funded by the Russian Science Foundation grant №.
22-11-00135, \url{https://rscf.ru/en/project/22-11-00135/}

\bibliographystyle{IEEEtran}
\bibliography{sample-base}

\end{document}